# Continuous variable entanglement measurement without phase locking

**Wang Dong (王东) \*\*, Huang Xianshan (黄仙山)**

*School of Mathematics and Physics, Anhui University of Technology, Maanshan 243000, China*

**Abstract**

A new simple entanglement measurement method is proposed for the bright EPR beams generated from a non-degenerate optical parametric amplifier operating at deamplification. Due to the output signal and idler modes are frequency degenerate and in phase, the needed phase shift of interference for the measurement of the correlated phase quadratures and anti-correlated amplitude quadratures can be accomplished by a quarter-wave plate and a half wave plate without separating the signal and idler beam. Therefore, phase locking and local oscillators are avoided.

In 1935，Einstein，Podolsky and Rosen (EPR) proposed a paradox in continuous variable measurements[1]. This refers to the concept of entanglement. In the end of 1980s, Reid and Drummond pointed out the quadrature amplitudes of optical fields output from an non-degenerate optical parametric oscillator (NOPO) could be propriate for demonstrating continuous variable EPR entanglement[2]. The signal and idler beams from an NOPO below or above threshold have correlated amplitude quadratures $\left(\left\langle \delta^2(X_2 - X_1) \right\rangle < 1\right)$ and anti-correalted phase quandratures[3,4] $\left(\left\langle \delta^2(Y_2 + Y_1) \right\rangle < 1\right)$. An NOPO, injecting bright subharmonic seed waves is named as an nondegenerate optical parametric amplifier (NOPA). Locking the injected subharmonic signal and harmonic pump field in phase or out of phase, an NOPA is in a state of parametric amplification or deamplification and can generate bright EPR beams with correlated amplitudes quadatures and anti-correlated phase quadratures or anti-corralated amplituede quadratures $\left(\left\langle \delta^2(X_2+X_1)<1\right\rangle\right)$ and correlated phase quadratures[5] $\left(\left\langle \delta^2(Y_2-Y_1)<1\right\rangle\right)$. Also, the two type EPR beams above can be obtained from two amplitude or two phase squeezed beams combined at a 50% beamsplitter with phase difference $\pi/2$.

In recent two decades, EPR beams led to a wide interest especially in quantum information[7] as a central role. A large number of protocols have been proposed and implemented, such as quantum teleportation[8], quantum dense coding[9], quantum key distributing[10], quantum swapping [11]et al.

For the measurement of EPR beams, the balanced homodyne detection is a common scheme for all type of EPR beams. Each of EPR beams needs a set of BHD system to measure the amplitude quadrature or phase quadrature[12]. In 2002, Jing Zhang and Kuichi Peng proposed a direct measurement scheme without local oscillators for the EPR beams with anticorrelation of amplitude quadratures and anticorrelation of phase quadratures. In 2010, Dong Wang proposed a similar direct measurement scheme for the EPR beams with correlation of amplitude quadratures and anti-correlation of phase quadratures[13]. For the frequency-nondegenerate twin beams, CV entanglement was experimentally with scanning a pair of tunable ring analysis cavities[14] or with two sets of unbalanced Mach-Zehnder interferometers[15], respectively. However, without exception, all of measurement schemes mentioned above needs locking the relative phase between EPR beams or EPR beams and local oscillators due to the thermal fluctuation. In this paper, a simpler scheme without phase locking is proposed for the signal and idler beams from an NOPA operating at deamplification, as shown in Fig.1.

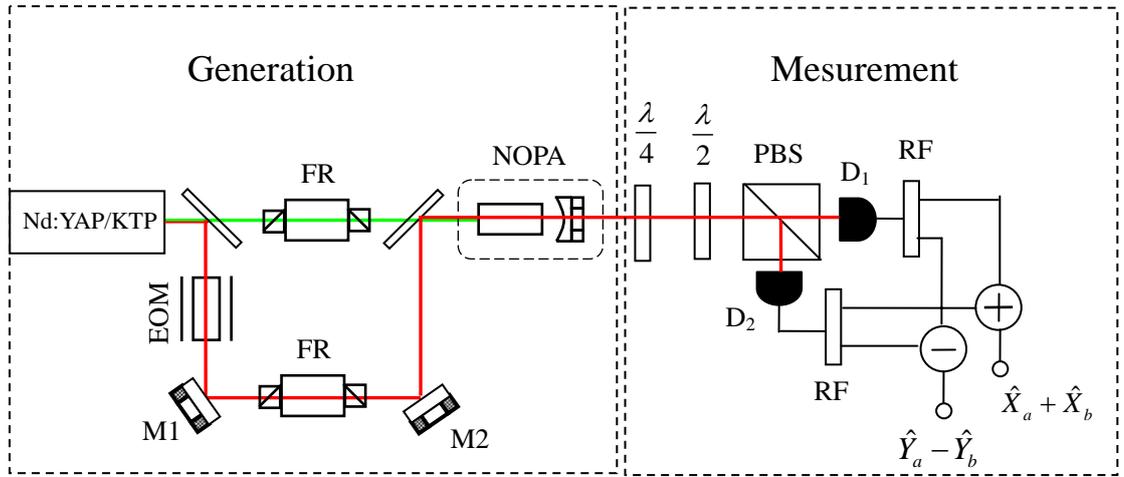

Fig.1 Generation Scheme for EPR beams with anti-correlated amplitude quadratures and correlated phase quadratures and direct measurement scheme without phase locking.

In Fig.1, An intracavity frequency-doubled laser serves as the light source of the pump wave and the seed wave for NOPA. The polarization of pump light is polarized along the b axis of the α-cut type-II crystal of the NOPA and the seed wave is polarized at 45° relative to the b axis of the KTP. After the seed wave is injected to NOPA, it is decomposed to signal and idler seed waves with identical intensity. By fine adjusting KTP temperature the signal and idler modes resonate simultaneously in NOPA. Once the double resonance is completed, the NOPA is locked on the frequency of the injected seed wave via a standard PDH technique[16]. The PZT on the output mirror of NOPA is controlled through the dispersion type error signal obtained by mixing mixing the modulation signal of (EOM) and the AC signal of the detector $D_0$. Hence the EPR beams output from NOPA are in phase. Without resonace, there is no certain phase relation.

In order to make NOPA operating in a state of parametric deamplification, it is required to lock the relative phase between the pump laser and injected seed wave to $(2n+1)\pi$ (n is integer). This can be accomplished by adjusting the PZT mounted on the mirror M2 and that is ok when the AC signal of the detector $D_0$ is max. In order to make this phase difference unchanged, a server loop is needed. First the output signal of a lock-in amplifier (LA) is impressed on the PZT of M1, then input the DC signal of $D_0$ to the LA and at last the output signal of (LA) can control the PZT of M2 to lock the $(2n+1)\pi$ phase difference. Therefore, the signal and idler beam output from NOPA have two properties below: first, their polarizations are perpendicular each other and they are in phase; second, their quadrature amplitudes are anticorrelated and quadrature phases are correlated.

Next, we discuss the measurement scheme shown in Fig.2. First, let the orthogonally-polarized EPR beams of signal mode $\hat{a}$ (s polarization) and idler mode $\hat{b}$ (p polarization) pass a $\lambda/4$ wave-plate which axes are parallel to the polarization directions of EPR beams. Note that the signal and idler mode are collinearly superimposed. So the relative phase is not sensitive to thermal fluctuation. Passing through the $\lambda/4$ plate, a phase difference of $\pi/2$ are made between the signal and idler light. So, the x-polarization mode can be denoted as $\hat{a}$ and y-polarization mode can be denoted as $i\hat{b}$.

Following is a half-wave plate which fast axes is 22.5° relative to the y direction.

It makes the signal and idle beams polarize in $\pm 45°$ direction as shown in Fig.2. The transmitted mode can be decomposed to the horizontal direction $\hat{p}$ and vertical direction $\hat{s}$ as follow:

$$\hat{p} = \frac{1}{\sqrt{2}}(\hat{a} - i\hat{b}) \qquad \hat{s} = \frac{1}{\sqrt{2}}(\hat{a} + i\hat{b}) \qquad (1)$$

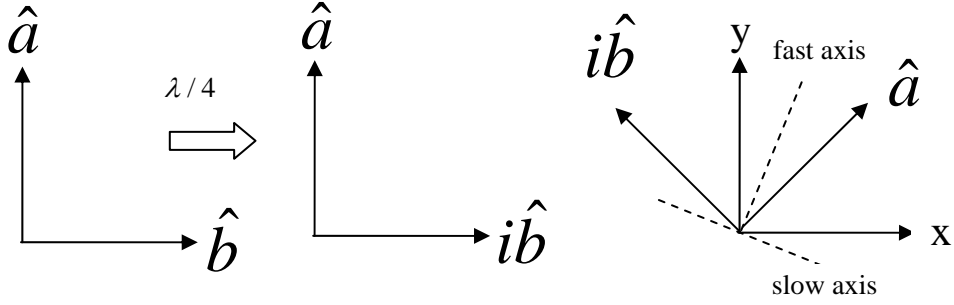

Fig.2　Phase and polarization of the EPR beams are transformed by a $\lambda/4$ wave-plate and a $\lambda/2$ wave-plate.

Then, The polarization-rotated beams are sent into a arm of a PBS shown in Fig.3. Here $\hat{p}$ is denoted as $\hat{a}_x$ and $\hat{s}$ is denoted as $\hat{a}_y$. With no input to the b arm, so both b polarizations are vacuum states.

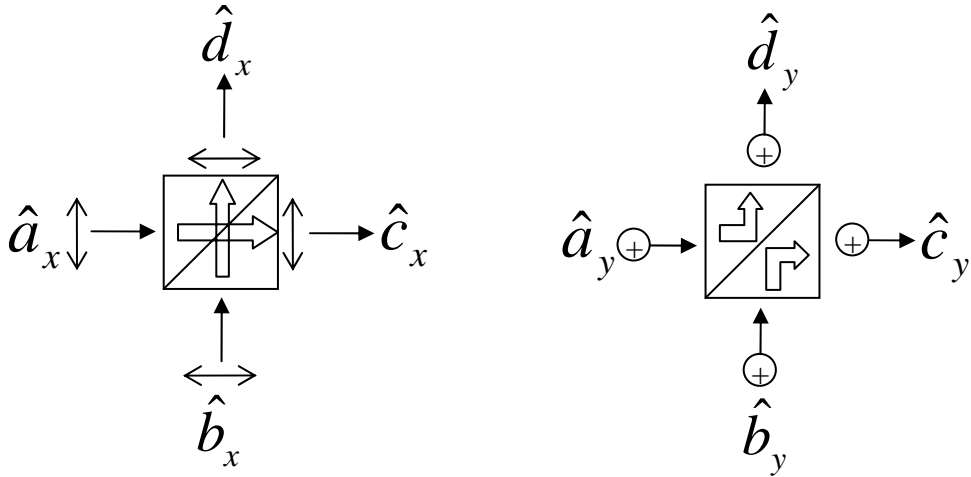

Fig.3 Polarizing beam splitter for input and output modes.

Photon numbers $\hat{n}_c = \hat{c}_x^+ \hat{c}_x$ and $\hat{n}_d = \hat{d}^+ \hat{d}$ in the two output ports of the interferometer are calculated by keeping the fluctuating contributions up to linear

terms:

$$\hat{n}_c = \frac{1}{2}\{2\alpha^2 + [\alpha\delta\hat{X}_a + \alpha\delta\hat{X}_b + \alpha(\delta\hat{Y}_b - \delta\hat{Y}_a)]\}$$
$$\hat{n}_d = \frac{1}{2}\{2\alpha^2 + [\alpha\delta\hat{X}_a + \alpha\delta\hat{X}_b + \alpha(\delta\hat{Y}_a - \delta\hat{Y}_b)]\}$$
(2)

where quadrature component X and Y are defined as $X = \hat{a}^+ + \hat{a}$ and $Y = i(\hat{a}^+ - \hat{a})$.

Each of the potocurrents is divided into two parts through the radio frequency splitters. The sum and difference of the divided photocurrents are

$$\delta^2 \hat{i}_+(\Omega) = \frac{\alpha^2}{2}[\delta^2(\hat{X}_a(\Omega) + \hat{X}_b(\Omega))]$$
$$\delta^2 \hat{i}_-(\Omega) = \frac{\alpha^2}{2}[\delta^2(\hat{Y}_a(\Omega) - \hat{Y}_b(\Omega))]$$
(3)

The sum of the correlation variances of the amplitude (X) and the phase (Y) quadratures of the twin beams is

$$\left\langle \delta^2(\frac{\hat{X}_a + \hat{X}_b}{\sqrt{2}}) \right\rangle + \left\langle \delta^2(\frac{\hat{Y}_a - \hat{Y}_b}{\sqrt{2}}) \right\rangle < 2$$
(4)

which demonstrates inseparability (entanglement) of EPR beams[17].

In summary, we proposed a simple direct measurement scheme for the EPR entanglement state without phase locking. According to the analysis above, the EPR beams must be anti-correlated for their amplitude quadratures and correlated for their phase quadratures. Except the generation scheme mentioned in this paper, the EPR beams could also be generated in self-phase-locked NOPA[18]. The presented scheme can be conveniently applied in some quantum state processing experiments such as quantum feedback[19], entanglement distillation[20] of non-Gaussian states et al.

In summary, we proposed a simple direct measurement scheme for the EPR

entanglement state without phase locking. According to the analysis above, the EPR beams must be anti-correlated for their amplitude quadratures and correlated for their phase quadratures. Except the generation scheme mentioned in this paper, the EPR beams could also be generated in self-phase-locked NOPA[18]. The presented scheme can be conveniently applied in some quantum state processing experiments such as quantum feedback[19], entanglement distillation[20] of non-Gaussian states et al.

This work supported by the Natural Science Research Key Program of Higher Education Institution of Anhui Province, China (Grant No. KJ2010A335).

**Email: wangdong@ahut.edu.cn